\documentclass[onecolumn]{emulateapj}

\newcommand{\sax}{{\it BeppoSAX}}

\newcommand{\xmm}{{\it XMM-Newton}}
\newcommand{\xte}{{\it RXTE}}
\newcommand{\integral}{{\it INTEGRAL}}
\newcommand{\chandra}{{\it Chandra}}
\newcommand{\source}{4U~1705--44}

\begin{document}

\title{The first detection of Compton Reflection in the Low-Mass X-ray Binary 4U~1705--44 with {\it INTEGRAL}{\thanks
{{\it INTEGRAL\/} is an ESA project with instruments and Science Data Centre funded by ESA member states (especially the PI countries: Denmark, France, Germany, Italy, Switzerland, Spain), Czech Republic and Poland, and with the participation of Russia and the USA.
}} and {\it BEPPOSAX}}


\author{Mariateresa  Fiocchi, Angela Bazzano, Pietro Ubertini}
\affil{Istituto di Astrofisica Spaziale e Fisica Cosmica di Roma (INAF)\\ Via Fosso del Cavaliere 100, Roma, I-00133, Italy} 
\author{Andrzej A. Zdziarski}
\affil{Centrum Astronomiczne im.\ M. Kopernika, Bartycka 18, 00-716 Warszawa, Poland}

\email{mariateresa.fiocchi@iasf-roma.inaf.it} 


\begin{abstract}
We present data from \integral\/ and \sax\/ satellites showing spectral state transitions of the neutron-star, atoll-type, low-mass X-ray binary \source. Its energy spectrum can be described as the sum of one or two blackbody components, a 6.4-keV Fe line, and a component due to thermal Comptonization. In addition, and for the first time in this source, we find a strong signature of Compton reflection, presumably due to illumination of the optically-thick accretion disk by the Comptonization spectrum. The two blackbody components, which the soft-state data require, presumably arise from both the disk and the neutron-star surface. The Comptonization probably takes place in a hot inner flow irradiated by some of the blackbody photons. The spectral transitions are shown to be associated with variations in the bolometric luminosity, most likely proportional to the accretion rate. Indipendentely from the spectral state, we also see changes in the temperature of the Comptonizing electrons and the strength of Compton reflection. 
\end{abstract}

\section{Introduction}

4U~1705--44 is a neutron-star low-mass X-ray  binary (LMXB) classified as an
atoll source (Hasinger \& van der Klis 1989). It also shows type-I X-ray bursts (Langmeier et al.\ 1987; Sztajno et al.\ 1985) and kHz quasi-periodic oscillations (Ford, van der Klis \& Kaaret 1998). The source shows variability on all time scales, from months down to milliseconds (Langmeier et al.\ 1987; Berger \& van der Klis 1998; Ford et al.\ 1998; Liu et al.\ 2001). On long time scales, it displays pronounced luminosity-related X-ray spectral changes between soft and hard states, as illustrated, e.g., using data from {\it Rossi X-ray Timing Explorer\/} (\xte) by Barret \& Olive (2002, hereafter BO02). The average spectrum of \source\ from the \integral/IBIS detector has been recently presented by Paizis et al.\ (2006).

As in other similar sources, the energy spectrum of \source\ has been described as the sum of a Comptonized component, a blackbody, and an emission line at $\sim 6.4$ keV. Using a \chandra/HETGS observation, Di Salvo et al.\ (2005) 
have shown that this Fe line is intrinsically broad (FWHM$\sim$1.2 keV), confirming White et al.\ (1986) and BO02, who reported before FWHM$\sim$1.1 keV. Similar broad emission lines at energies in the range of 6.4--6.7 keV are often observed in the spectra of LMXBs, both in systems containing black holes \cite{miller02} and old neutron stars \cite{ba01asr}. 

Compton reflection of X-rays, a process studied in detail in the last few decades (e.g., Lightman \& Rybicki 1979; White, Lightman \& Zdziarski 1988; Lightman \& White 1988; Matt, Perola \& Piro 1991; Magdziarz \& Zdziarski 1995), takes place when X-rays and $\gamma$-rays interact with a cold medium. The X-ray spectrum of Compton reflection has a characteristic shape resulting from photoelectric absorption at low energies and Compton scattering (including its recoil) at higher energies, with a broad hump around 30 keV. Such a component has never been reported for \source.

In this paper, we present results of a study of the spectral variability of 4U~1705--44 during state transitions observed recently by \integral\/ during its monitoring of the Galactic Centre region, and earlier by \sax. Data analysis is described in \S\ \ref{observations}. The spectral variability and its analysis are given in \S\ \ref{analysis}. Finally, \S\ \ref{discussion} gives discussion and interpretation of our results. 

\begin{table}
\caption{The log of \sax\/ and \integral\/ observations.}
\label{jou}
\scriptsize
\centering
\vspace{0.1cm}
\begin{tabular}{lccccccc}
\hline\hline
& & & & & & &\\
\multicolumn{8}{c}{\it{BeppoSAX}}\\
& {Start Date}&&{Exposure [ks]}&&&{Count rate [s$^{-1}$]$^{a}$}&\\

&&LECS&MECS&PDS&LECS&MECS&PDS\\
&& & & &[0.5--3.8 keV]&[1.5--10 keV]&[20--100 keV]\\
& & & & & & &\\

{$1^{\rm st}$ epoch} &2000-08-20 &21 &44 &20 &$3.88\pm 0.01$  &$27.25\pm 0.03$ &$0.90\pm0.04$\\
{$2^{\rm nd}$ epoch} &2000-10-03 &16 &48 &20 &$0.48\pm 0.01$  &$4.46\pm 0.02$  &$7.86\pm0.05$\\
& & & & & & &\\
\hline
& & & & & & &\\
\multicolumn{8}{c}{\it{INTEGRAL}}\\

&&{Start Date}&\multicolumn{2}{r}{Exposure [ks]}&\multicolumn{2}{c}{Count rate [s$^{-1}$]}&\\
&& &JEM-X&IBIS&JEM-X&IBIS&\\
&&& & &[4.5--15keV]&[20--150 keV]&\\

& & & & & &&\\
&{$3^{\rm rd}$ epoch}  &2003-02-02  &106  &394   &$11.63\pm 0.04$ &$1.60\pm0.05$&\\
&{$4^{\rm th}$ epoch} &2003-08-09  &17  &93 &$8.6\pm 0.1$ &$12.6\pm 0.1$&\\
&& & & & &&\\
\hline
\hline
\end{tabular}

{$^{a}$ The MECS count rates correspond to the sum of the MECS2 and MECS3 units.}
\end{table}

\section{Observations and Data Analysis}
\label{observations}

Table \ref{jou} gives the log of \integral\/ and \sax\/ observations of the source performed with LECS, MECS, and PDS and JEM-X and IBIS on board of \sax\/ and \integral\/ satellites, respectively. Unfortunately the HP-GSPC data, covering the interesting energy band 3-120 keV are not available, since this instrument was not working properly.
Moreover, we can not use the IBIS/PICSIT data being the \source\/ emission below the spectral sensitivity threshold neither the SPI data because of its angular resolution of $\sim$ 2 $\degr$ is not adeguate to resolve the emission of this source.
\sax\/ observed \source\ twice, in 2000 August and October. 
The LECS, MECS and PDS event files and spectra, available from the ASI Scientific Data Center(ASDC), were generated with the Supervised Standard Science Analysis (Fiore, Guainazzi \& Grandi 1999). Both LECS and MECS spectra were accumulated in circular regions of $8\arcmin$ radius. The PDS spectra were obtained with the background rejection method based on fixed rise time thresholds. Publicly available matrices were used for all the instruments. Spectral data were then binned using the template files distributed by ASDC, in order to ensure the applicability of the $\chi^2$ statistic and an adequate sampling of the spectral resolution of each instrument. Spectral fits were performed in the following energy bands: 0.5--3.8 keV for the LECS, 1.5--10.0 keV for the MECS and 20--150 keV for the PDS.

\begin{figure}
\centering
\includegraphics*[angle=-90,scale=0.35]{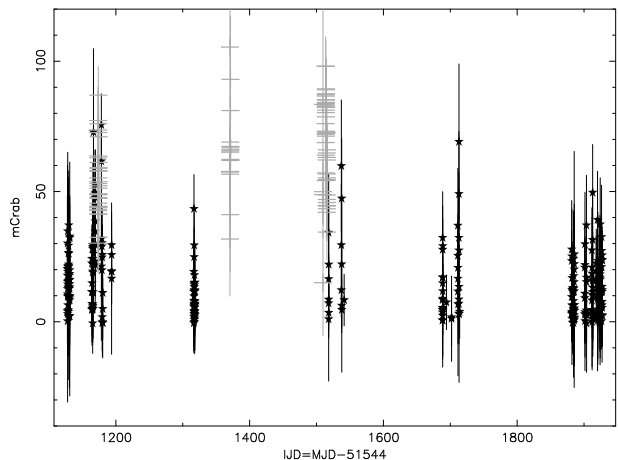}
   \caption{The IBIS light curve in the 40--60 keV energy band. The stars denote data corresponding to relatively steady emission, and the crosses denote data with the average flux $>50\%$ higher than in the average steady level. }
\label{lci}
\end{figure}

The analyzed \integral\/ (Winkler et al.\ 2003) data consist of all public observations in which 4U 1705--44 was within the field of view (FOV) of high-energy detectors. The observations are divided into uninterrupted 2000-s intervals, the so-called science windows (SCW), within which the light curves and spectra are extracted individually. Broad-band spectra, $\sim$5--150 keV, are obtained using data from the high-energy instruments, JEM-X (Lund et al.\ 2003) and IBIS (Ubertini et al.\ 2003). The IBIS and JEM-X data have been processed using the Off-line Scientific Analysis (OSA v.\ 5.0) software released by the \integral\/ Science Data Centre (ISDC, Courvoisier et al.\ 2003). While IBIS  provides a large FOV ($\sim 30\degr$), JEM-X has a narrower FOV ($<10\degr$), thus providing only a partial overlap with IBIS. Data from the Fully Coded field of view only for both instrument have been used 
($4.5\degr$ and $2.5\degr$ for IBIS and JEM-X, respectively). Spectra were extracted only if signal-to-noise ratio was $>10$.

\section{Results}
\label{analysis}

In our spectral analysis of the steady emission, we aim to determine which parameters drive the changes between the spectral states. To this aim, we fit the data with several physical models, while trying to minimize the number of free parameters. Each time a new component is added to the model, the F-test (Bevington \& Robinson 1992) is performed (but note caveats to its validity in certain cases, Protassov et al.\ 2002). During spectral fitting, we allow the relative normalization to be free with respect to the MECS and IBIS data, for \sax\/ and \integral, respectively. We use XSPEC v.\ 11.3.1.

Fig.\ \ref{lci} shows the IBIS light curve in the 40--60 keV energy band,
where we expect main spectral changes for atoll sources. As it can be seen, the source flux has been observed to increase more than $50\%$ on three occasions, during MJD 52712--52718, 52914--52916 and 53044--53060, as showed in Fig.\ \ref{lci} as crosses. However, we find substantial spectral changes only during the last two intervals. 

We study the spectral behavior separately in four epochs consisting of the following data:
\begin{itemize}
\item
{$1^{\rm st}$ epoch}. The \sax\/ observation performed on MJD 51777, in the soft/high state.
\item
{$2^{\rm rd}$ epoch}. The \sax\/ observation performed on MJD 51820, in the hard/low state.
\item
{$3^{\rm rd}$ epoch}. The JEM-X and IBIS observations between MJD 52671 and 53471 but excluding the observations of MJD 52914--52916 and 53044--53060. The source was in the soft/high state.
\item
{$4^{\rm th}$ epoch}. The JEM-X and IBIS observations of MJD 52914--52916 and 53044--53060, in the hard/low state.
\end{itemize}

The simplest model which provides a reasonable fit to the \sax\/ data consists of the sum of a thermal Comptonization component, modeled in XSPEC by
{\scriptsize{COMPTT}} (Titarchuk 1994; spherical geometry was assumed),
a soft component modeled by a single temperature blackbody, and a Gaussian Fe line at 6.4 keV with the width, $\sigma_{\rm Fe}$. This spectrum is absorbed by a column density, $N_{\rm H}$. The main free parameters of this model are the blackbody temperatures, $T_{\rm bb}$, the temperature, $T_{\rm e}$, and the optical depth, $\tau$, of the thermal electrons, and the temperature of the blackbody seed photons irradiating the thermal plasma, $T_0$. 

The $N_{\rm H}$ fitted to the \sax\/ data is consistent with that given by 
Langmeier et al.\ (1987), (0.7--$1.9)\times 10^{22}$ cm$^{-3}$. Initially, the Fe line centroid energy was let free; however, our fits show it to be always consistent with the neutral-Fe value of 6.4 keV, and hence it has been fixed at that value hereafter. 

Adding Compton reflection to the fit, using the XSPEC model {\scriptsize{REFLECT}} (Magdziarz \& Zdziarski 1995), resulted in a substantial fit improvement for epoch 2, reducing $\chi^2/$d.o.f.\ from $510/414$ to $500/413$. This corresponds to the low F-test chance probability of $4\times 10^{-3}$. The inclination angle, $i$, of the reflecting medium has been kept at $63\degr$ ($\cos i=0.45$), which is consistent with the allowed range found by Di Salvo et al.\ (2005) of $i=55$--$84\degr$ from \chandra\/ observation. The strength of reflection is measured by the solid angle, $\Omega$, subtended by the reflection, with $\Omega=2\pi$ corresponding to reflection from a slab completely covering the disk. The blackbody temperatures are $kT_{\rm bb}=0.58$ keV and 0.65 keV for the soft and hard state, respectively. With this model, the main parameters agree with ones reported by BO02, with only exception of the blackbody temperature, which is now slightly higher in the hard  state.

The \integral\/ data have been fitted with a model similar to that used for the \sax\/ data (but initially without Compton reflection). However, since both the energy resolution of JEM-X, $\Delta E/E \sim 1$ (FWHM) at 6.4 keV is not adequate to study the Fe line properties and the usable JEM-X data are for $\ga 5$ keV only, we do not include the line in the present model. The seed soft temperature and the column density are now frozen at their respective values obtained from the \sax\/ data. This model gives adequate values of $\chi^{2}/$d.o.f.\ for the data sets of epochs 3 and 4, 128/111 and 129/118, respectively. However, we find the fitted blackbody temperature for the hard state, epoch 4, to be extremely high, 6.1 keV, which is clearly unphysical. 
The corresponding component, yielding a hump around $\sim$20--30 keV, 
is very well reproduced by Compton reflection. Similary to the case of the 
\sax\/ data, we add the Compton reflection to the fit. Also, since the \integral\/ energy range does not allow for a reliable determination of a low-energy blackbody, we fix its temperature at the value obtained from the 
hard state \sax\/ data.

\clearpage
\begin{table}
\centering
\tiny
\caption{The fit results. The symbols H and S refer to the soft/high and hard/low state, respectively.
'f' denotes a fixed parameter. }
\label{fit}
\vspace{0.1cm}
\begin{tabular}{c@{\ \ }c@{\ \ }c@{\ \ }c@{\ \ }c@{\ \ }c@{\ \ }c@{\ \ }c@{\ \ }c@{\ \ }c@{\ \ }c@{\ \ }c@{\ }c@{\ }c@{\ }c}
\hline\hline
& & & & & & & & & &&&&\\
Epoch, &{$N_{\rm H}$} &{$kT_{\rm bb1}$} &{$kT_{\rm {\rm bb2}}$} & $kT_{0}$  &{$kT_{\rm e}$} &$\tau$ &$\Omega/2\pi$ &$\sigma_{\rm Fe}$& EW  &$R_{\rm bb1}$ &$R_{{\rm bb2}}$&$R_{\rm seed}$& $n_{\rm comptt}$ & $\chi^2$/d.o.f. \\
State &$10^{22}$cm$^{-2}$ &keV &keV&keV &keV &&& keV &eV  &     km&km  &km     &$10^{-2}$ &\\
& & & & & & & & & &&&&\\
\hline

& & & & & & & & & &&&&\\
& & & & & & & & & &&&&\\
1, S&$1.37\pm0.03$  &$0.52^{+0.03}_{-0.01}$&$1.64\pm0.02$  &$=kT_{\rm bb1}$ & $18.8_{-0.3}^{+0.7}$  &$0.40\pm0.05$&$1.1\pm0.2$ & $<0.3$ & $8^{+10}_{-2}$  &$<20$ &
$3.3\pm0.2$&46& $4.9^{+0.5}_{-1.0}$ & 567/422\\

& & & & & & & & & &&&&\\

& & & & & & & & & &&&&\\
2, H &$1.2_{-0.1}^{+0.1}$ &$0.65_{-0.06}^{+0.05}$&&$1.14_{-0.08}^{+0.04}$   &$22_{-1}^{+4}$ &$1.5_{-0.6}^{+0.3}$        &$0.2_{-0.1}^{+0.2}$  & $0.8^{+0.3}_{-0.3}$ & $189_{-39}^{+20}$  & $8.2_{-0.1}^{+0.1}$&&7 &$0.7\pm0.1$ & 500/413\\
& & & & & & & & & &&&&\\

& & & & & & & & & &&&&\\
3, S & 1.37f&0.52f&$2.17\pm0.04$
&$=kT_{\rm bb1}$ &$19.5\pm0.8$ &$0.27\pm0.03$ &$<0.6$ & & &$<26$& $2.4\pm0.2$ &80&$16\pm2$&120/109\\
& & & & & & & & & &&&&\\

& & & & & & & & & &&&&\\
4, H &1.3f &0.65f& &1.14f & $49\pm3$  &$0.3\pm0.1$&$1.4_{-0.3}^{+0.5}$  & & &$<17$&&8&$0.5\pm0.2$ &118/116\\
& & & & & & & & & &&&&\\
\hline
\hline
\end{tabular}\\
\end{table}

\begin{figure}
\centering
\includegraphics[angle=-90,width=7.5cm]{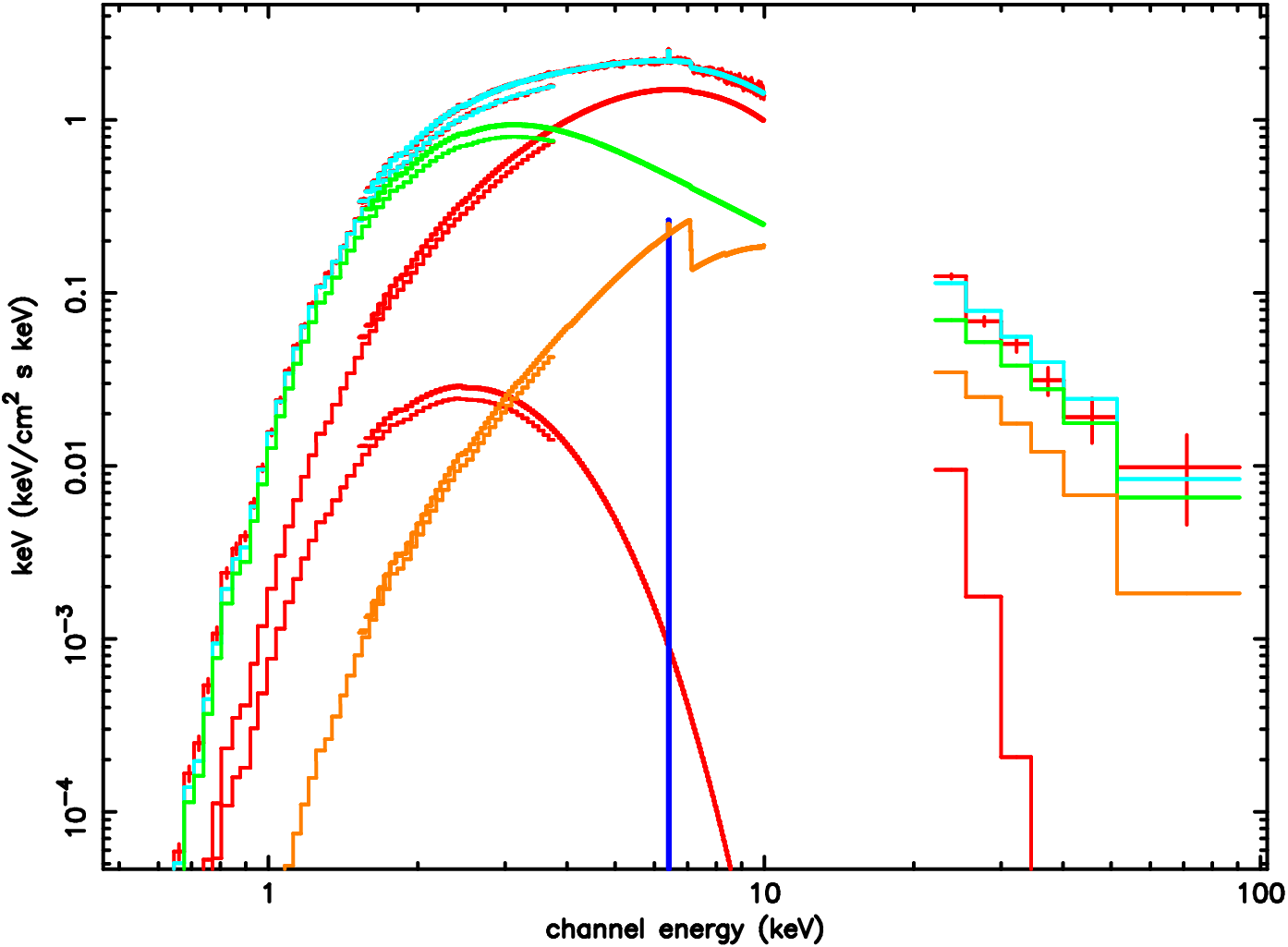}
\includegraphics[angle=-90,width=7.5cm]{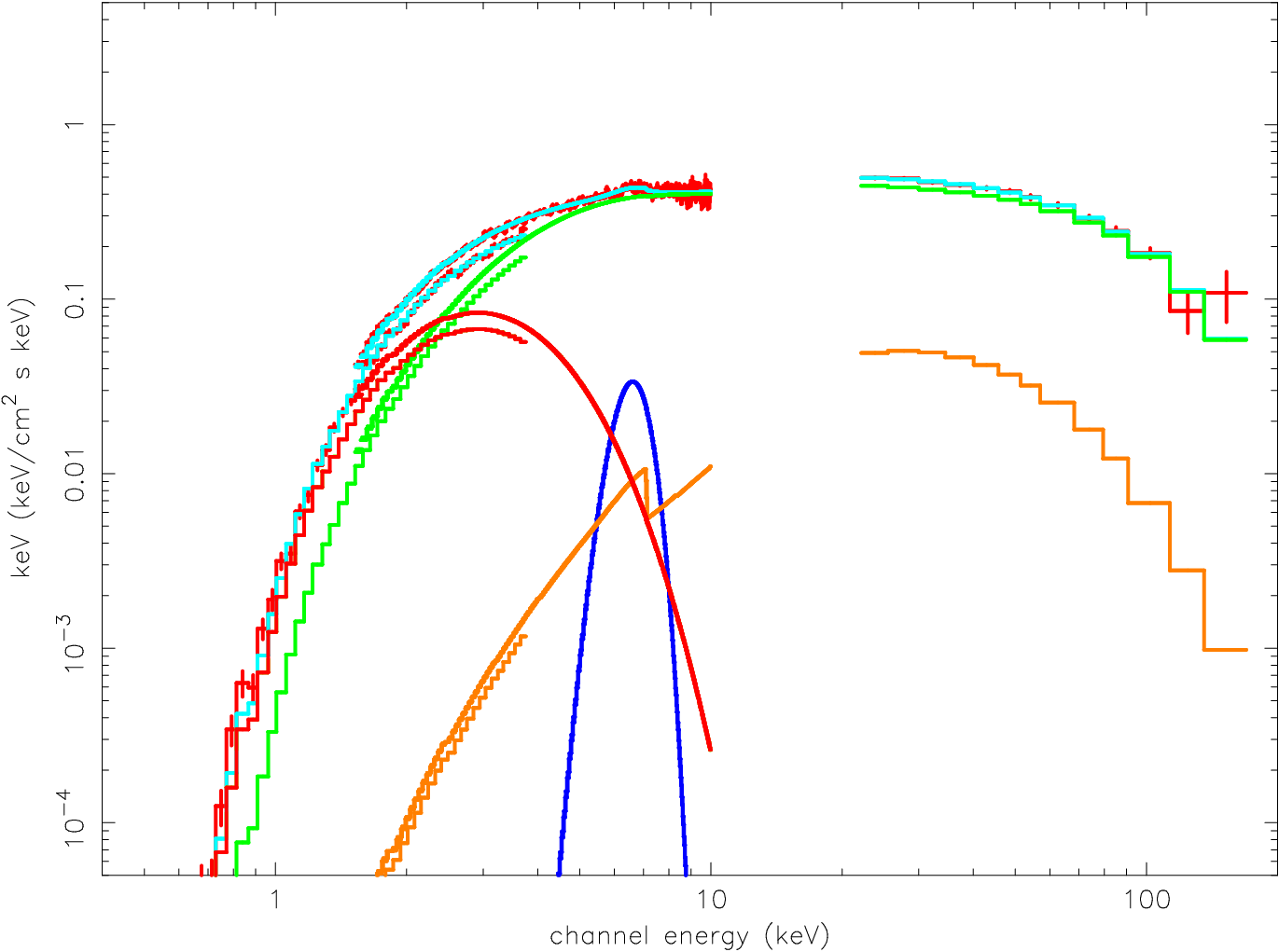}
   \caption{The spectra of epochs 1 (soft state, left) and 2 (hard state, right) observed by \sax\/, shown together with the total model and its components. Left: the blackbody, Comptonization and the Fe line components are shown in red, green and blue, respectively. Right: the blackbody, Comptonization, reflection and Fe line are shown in red, green, orange and blue respectively.
}
\label{sax1}
\end{figure}

\begin{figure}
\centering
\includegraphics[angle=-90,width=7.5cm]{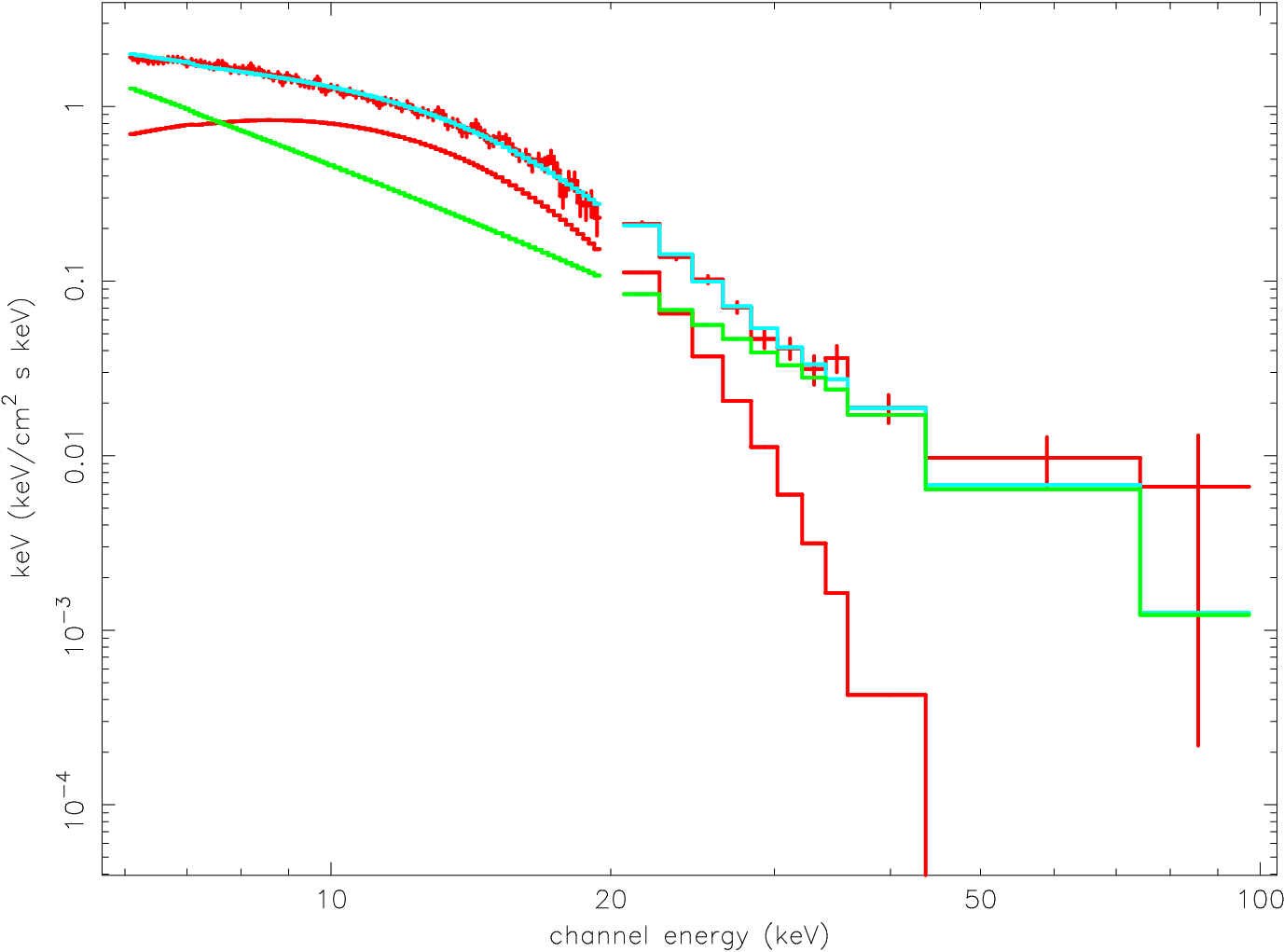}
\includegraphics[angle=-90,width=7.5cm]{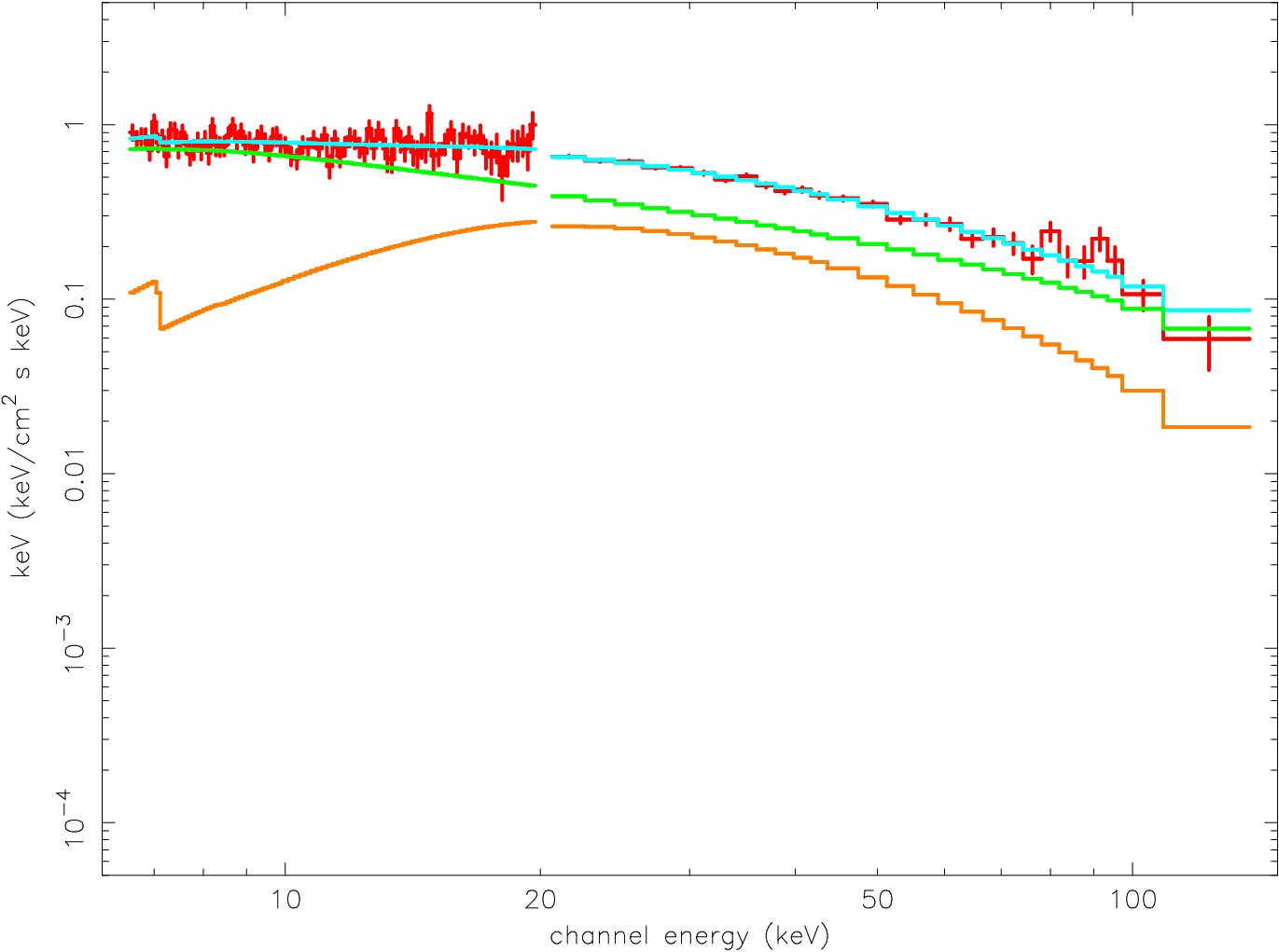}
\caption{The spectra of epochs 3 (soft state, left) and 4 (hard state, right) observed by \integral, shown together with the total model and its components. Left: the blackbody and Comptonization components are shown in red and green, respectively. Right: Comptonization and the reflection component is shown in green and orange respectively.}
\label{int1}
\end{figure}

On the other hand, this procedure still does not yield satisfactory results for the epoch 3. By fixing $kT_{\rm bb}$ at the value obtained from the \sax\/ soft state data, we obtain an unacceptable $\chi^2/{\rm d.o.f.}\sim 2$. A higher value, $kT_{\rm bb}\sim 2.5$ keV is required by the data. Therefore, we now fit both the \sax\/ and \integral\/ soft states with two blackbody components and set $kT_{\rm 0}$ equal to the lower value of their temperatures, $kT_{\rm bb1}$. This reduces $\chi^{2}/$d.o.f.\ from 578/423 to 567/422 for the epoch 1 and from 195/110 to 120/109 for the epoch 3, with the low corresponding F-test chance probabilities of $4\times 10^{-3}$ and $4\times 10^{-13}$, respectively. Using this model, the reflection component becomes statistically highly significant for the epoch 1, reducing $\chi^{2}/$d.o.f.\ from 632/423 to 567/422, which corresponds to the F-test chance probability of $1.3\times 10^{-11}$. Spectral fit results are given in Table \ref{fit} and the spectra are shown in Fig.\ \ref{sax1} and \ref{int1} for the \sax\/ and \integral\/ data, respectively.

\begin{figure}
\centering
\includegraphics[angle=-90,width=7.5cm]{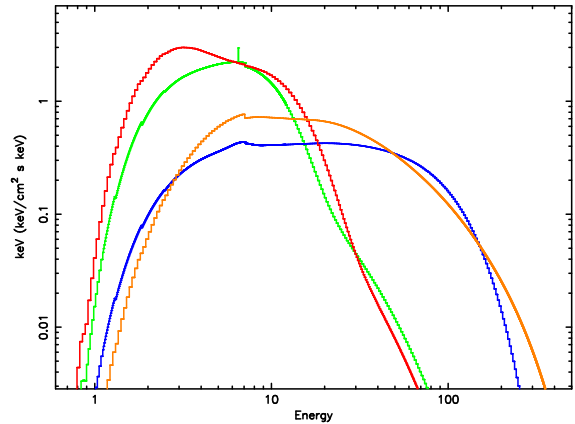}
\caption{Comparison of the models for the four observations. Epoch 1, soft state: green; epoch 2, hard state: orange; epoch 3, soft state: red; epoch 4, hard state: blue.
}
\label{models}
\end{figure}

Summarizing the above results, the energy spectra of 4U~1705--44 can be 
fitted by the sum of one (in the hard state) or two (in soft state) blackbody components, a thermal-Comptonization component and a 6.4 keV Fe line, in agreement with the results of BO02. We find two new results. First, two soft blackbody components are required by the data in the soft state. Second, we find strong signature of Compton reflection. The four models are compared in Fig.\ \ref{models}. We clearly see the distinct difference between the hard and soft states.
 
The bolometric model luminosities (obtained in the 0.1--200 keV range) 
are $L \simeq 4.6 \times 10^{37}$ erg s$^{-1}$ and $\simeq 8.1 \times 10^{37}$ erg s$^{-1}$, for the soft-state epochs 1 and 3, respectively, assuming the distance of $D=7.4$ kpc \cite{ha95}. They correspond to 22\% and 39\%, respectively, of the Eddington luminosity, $L_{\rm E}$, for $M=1.4{\rm M}_\sun$ 
and the standard abundances. Most, $\sim 80\%$, of the source flux, 
is radiated below 20 keV.

The bolometric model luminosities are $L \simeq 1.5 \times 10^{37}$ erg s$^{-1}$ 
and $\simeq 2.1 \times 10^{37}$ erg s$^{-1}$, for the hard-state epochs 2 and 4, respectively. They correspond to 7\% and 10\% of $L_{\rm E}$. They are both lower than those in the soft state, in agreement with the usual ranking of the luminosity of these two states, in particular in atoll sources (e.g., Hasinger \& van der Klis 1989; van der Klis 2000; Gierli\'nski \& Done 2002). The electron temperatures are now substantially higher than in the soft state, 
$kT_{\rm e}\sim 20$--50 keV, and the Comptonization component extends above $\sim 100$ keV. 

The seed-photon temperatures are consistent with values typical for neutron-star LMXBs, $kT_{0}\sim 0.3$--1.5 keV \cite{Oo01}. The blackbody temperatures of the second, hotter, component are relatively high during the soft states. We note that similar high temperatures have been found by BO02 in their soft-state \xte\/ observations. Also, they found that the blackbody temperature increases with the bolometric luminosity. We confirm this trend for the \integral\/ soft state, which has $L$ higher than those of BO02, and for which the both blackbody temperatures are higher than the values reported by those authors. On the other hand, no significant variations of either $kT_{0}$ or $N_{\rm H}$ have been detected, with our values consistent with those of BO02 and Langmeier et al.\ (1987).

We have also estimated the sizes of the regions emitting the seed photons for Comptonization, assuming their emission as blackbody (following in 't Zand et al.\ 1999). We obtain $R_{\rm seed}\simeq 8.8\,{\rm km}\, (L_{\rm C}/10^{37}\,{\rm erg\, s}^{-1})^{1/2} (1+y)^{-1/2} (kT_0/1\,{\rm keV})^{-2}$, where $L_{\rm C}$ is the luminosity in the Comptonization component. Here, we have estimated the amplification factor of the seed luminosity by the Comptonization as $(1+y)$, where $y$ is the Comptonization parameter, $y=4kT_{\rm e}\max(\tau,\tau^2)/m_{\rm e} c^2$. The obtained values are given in Table \ref{fit}, and are in the range between 7 and 80 km. 

\section{Discussion}
\label{discussion}

As one of our main results, we have found two blackbody components required by the data in the soft state. The hotter one is likely originating from a part of the neutron star surface or the innermost part of the disk, with $kT_{\rm bb}\simeq 1.6$--2.2 keV and the radius of few km. The colder one, $kT_{\rm bb}\simeq 0.5$ keV, is likely produced farther away from the neutron star, $R_{\rm bb}\la 80$ km. Our fitted values of the radius in the unscattered blackbody component are much less than the inferred radius of the blackbody photons scattered by a surrounding hot plasma. This is an indication that most of those colder blackbody photons undergo Compton scattering. In the hard state, only one blackbody component at a low temperature, $kT_{\rm bb}\simeq 0.6$--0.7 keV, is present. Its emission region radius is constrained mostly by the \sax\/ data to $\sim 8$ km, which most likely corresponds to the stellar surface (cf.\ Olive, Barret \& Gierli\'nski 2003).

The Comptonization component may arise from a corona above the disk and/or 
between the disk and the stellar surface. The obtained values of the electron temperature are typical for neutron-star LMXBs, but lower than those seen in the hard state of black-hole binaries, $kT_{\rm e}\sim 100$ keV (Gierli\'nski et al.\ 1997; Di Salvo et al.\ 2001; Zdziarski \& Gierli\'nski 2004).
We also find the electron temperature to be higher in the hard state.

Although Barret et al. 2001 observed spectral state transitions
without significant variations in the total X-ray luminosity,
we found that they are accompanied by changes in luminosity,
indicating they are driven by variability of the accretion rate
that is lower in the hard state, and higher in the soft state.
For an accretion efficiency of $\eta = 0.2$ (corresponding, e.g., to $M_{\rm NS}=1.4 {\rm M}_{\sun}$ and $R_{\rm NS}=10$ km) and using our model luminosities, $L$,
we find
$\dot{M}_1\simeq 4.1 \times 10^{-9} {\rm M}_{\sun}$ yr$^{-1}$,
$\dot{M}_2\simeq 1.3 \times 10^{-9} {\rm M}_{\sun}$ yr$^{-1}$,
$\dot{M}_3\simeq 7.1 \times 10^{-9} {\rm M}_{\sun}$ yr$^{-1}$ and
$\dot{M}_4\simeq 1.9 \times 10^{-9} {\rm M}_{\sun}$ yr$^{-1}$.
They are apparently accompanied by changes the geometry of the flow,
and the relative contribution of the blackbody-like and Comptonization components. In the soft state, the blackbody components are much stronger than in the hard state.

The spectral evolution of atoll sources could be explained in the framework of a model consisting of a truncated accretion disk with a hot inner flow (e.g., BO02; Olive et al.\ 2003). The truncation radius of the disk is the critical parameter of this model, as the disk represents a major source of cooling for the Comptonizing plasma. In the hard states, accretion probably assumes the form of a truncated outer accretion disk and a hot inner flow, joining the disk and the stellar surface.

In the two \sax\/ data sets, we have detected a line emission which we interpret as K-shell fluorescent emission of low ionization states of Fe. For the first time in the source, we have detected strong signatures of Compton reflection. Earlier, Compton reflection was detected in a number of neutron-star LMXBs, e.g., 4U 1608--522 (Zdziarski, Lubi\'nski \& Smith 1999; Gierli{\'n}ski \& Done 2002), GS 1826--238 (Zdziarski et al.\ 1999; Barret et al.\ 2000), SLX 1735--269 (Barret et al.\ 2000), 4U 0614+09 (Piraino, Santangelo \& Kaaret 2000), Cyg X-2 (Done, \.Zycki \& Smith 2002), SAX J1808.4--3658 (Gierli\'nski, Done \& Barret 2002), 4U 1820--303 (Ballantyne \& Strohmayer 2004; Tarana et al.\ 2006), and GX 1+4 (Rea et al.\ 2005). 
We note, however, that we find a strong Fe line in the epoch 2,
when the reflection is weak. This indicates that the line flux
and the reflection strength may not be always correlated,
which could indicate these components originate from different region.
Although \sax\/ data strongly support this indication, better quality data
at low energy are necessary to address this behavior.

The large width of the Fe line detected in the epoch 2 could be due to 
either blending of several ionization states, relativistic effects in the accretion disk, or Compton broadening in a surrounding corona. Unfortunately, our data do not allow us to measure the detailed profile of the line, that would discriminate between these hypotheses. We note that the latter is also the case for the \chandra\/ data \citep{disalvo05}, and that the large effective areas of \xmm\/ would probably give sufficient statistics to determinate the origin of the line width. 

\begin{acknowledgments}
We acknowledge the ASI financial/programmatic support via contracts ASI-IR 
046/04. AAZ has been supported by the Polish grants 1P03D01827, 1P03D01128 and 4T12E04727. Special thanks are due to M. Federici for supervising the \integral\/ data archive.
\end{acknowledgments}

\end{document}